\documentclass[conference]{IEEEtran}
\IEEEoverridecommandlockouts
\makeatletter
\newcommand{\ssymbol}[1]{^{\@fnsymbol{#1}}}
\makeatother

\usepackage{cite}
\usepackage{amsmath,amssymb,amsfonts}
\usepackage{algorithmic}
\usepackage{graphicx}
\usepackage{textcomp}
\usepackage{xcolor}
\usepackage[hyphens]{url}
\def\BibTeX{{\rm B\kern-.05em{\sc i\kern-.025em b}\kern-.08em
    T\kern-.1667em\lower.7ex\hbox{E}\kern-.125emX}}
\begin{document}

\title{TSCH Evaluation under heterogeneous Mobile Scenarios\\
}

\author{\IEEEauthorblockN{Charalampos Orfanidis$\ssymbol{2}\ $, Atis Elsts$ \ssymbol{3}\ $, Paul Pop$\ssymbol{2}\ $, Xenofon Fafoutis$\ssymbol{2}\ $}
\IEEEauthorblockA{$\ssymbol{2}\ $Department of Applied Mathematics and Computer Science, Technical University of Denmark \\
$ \ssymbol{3}\ $Institute of Electronics and Computer Science (EDI)}
Email: {\{chaorf, paupo, xefa\}@dtu.dk}, atis.elsts@edi.lv}

\maketitle

\begin{abstract}
Time Slotted Channel Hopping (TSCH) is a medium access protocol defined in the IEEE 802.15.4 standard. It has been demonstrated to be one of the most reliable options when it comes to industrial applications. TSCH offers a degree of large flexibility and can be tailored to the requirements of specific applications. Several performance aspects of TSCH have been investigated so far, such as the energy consumption, the reliability, scalability and many more. However, mobility in TSCH networks remains an aspect that has not been thoroughly explored. In this paper we examine how TSCH performs under mobility situations. We define two mobile scenarios: one where autonomous agriculture vehicles move on a predefined trail, and a warehouse logistics scenario, where autonomous robots/vehicles and workers move randomly. We examine how different TSCH scheduling approaches perform on these mobility patterns and when different number of nodes are operating. The results show that the current TSCH scheduling approaches are not able to handle mobile scenarios efficiently. Moreover, the results provide insights on how TSCH scheduling can be improved for mobile applications. 
\end{abstract}

\begin{IEEEkeywords}
IoT, TSCH, Mobility, Reliability, Robustness
\end{IEEEkeywords}

\section{Introduction}
\label{sec:Introduction}

Time Slotted Channel Hopping (TSCH) is a Medium Access Control (MAC) protocol for Internet of Things (IoT) applications, which has been used broadly in industry since it can offer high level of reliability and robustness \cite{Duquennoy2017}. TSCH is implemented as a channel hopping mechanism to avoid interference and multi-path fading which Low-power and Lossy Networks (LLN) are prone to. Furthermore, all devices in a TSCH network are synchronized to a common time source, the coordinator node. Essentially all the nodes are synchronized (with a tolerated offset) and a schedule is calculated. The purpose of the schedule is to dictate which node is supposed to perform an action (receive, transmit, sleep), on which channel and at which point in time. 

The modular nature of TSCH has enabled high flexibility and therefore, there are numerous application scenarios which have utilized the TSCH mechanism. Industrial monitoring \cite{WirelessHART}, environmental monitoring\cite{Watteyne2016}, smart home for healthcare\cite{Elsts18}, smart buildings \cite{BRUNLAGUNA201883} and many more. Since there are different requirements in every application scenario, there are several modified versions of TSCH in this regard. One common way to tailor TSCH based on the application requirements is to have a different approach on the scheduling scheme. Thus, there is a rich scientific literature describing different scheduling approaches for TSCH which can be summarized in three main categories: centralized, distributed, autonomous. In centralized scheduling \cite{WirelessHART} a node is responsible to calculate the schedule and distribute it to the rest nodes. In distributed scheduling approaches, the schedule is constructed based on the relations between neighboring nodes \cite{rfc4180}. In the autonomous approaches each node is constructing its own schedule mostly based on routing information that are present on the node already \cite{Duquennoy2015}.

Several aspects of TSCH have been explored such as energy consumption \cite{Vilajosana2014}, scalability \cite{Hamideh2019, Duquennoy2015}, and latency\cite{Guglielmo2016}. However it is still unclear how a TSCH network is performing under a mobile scenario since that topic is not explored thoroughly in the scientific literature. The mobility feature in IoT is emerging as application scenarios like autonomous robots and vehicles \cite{Bigi2019} operate in remote areas in some cases, where there is no access on cellular network. Furthermore other applications like warehouse logistics \cite{Xiulong2018} for instance, where a synergy of moving machinery, autonomous robots/vehicles and workers might require low power mobile communication to fulfil the application requirements. 

In this paper, the main research question we try to answer is how reliable are the current TSCH schedulers under different mobile scenarios. We consider two mobile scenarios and based on simulations we evaluate three different schedulers to examine the reliability of TSCH. Specifically we evaluate how mobility affects its reliability in terms of Packet Received Ratio (PRR), downtime, and the initial joining time. In addition we consider how different number of nodes in a network architecture would impact the network reliability. The results illustrate that the current TSCH approaches do not operate efficiently in that cases and we provide some proposals towards improving them. As a main evaluation tool we use Cooja \cite{cooja} simulator. 

The rest of the paper is organized as follows: Section \ref{sec:Background} describes basic principles of TSCH and the schedulers used in this paper, Section \ref{sec:RelatedWork} lists the state of the art and how it relates to the presented approach. Section \ref{sec:MobilityPatternsAndApplications} illustrates the technical details and motivation behind the selected mobility patterns and the representing applications, Section \ref{sec:Evaluation} presents the evaluation method and the obtained results and Section \ref{sec:Conclusion} discusses future work and concludes the paper.

\section{Background}
\label{sec:Background}

This section provides a technical overview of TSCH protocol and the TSCH schedulers which are used in the evaluation part. TSCH is included in the IEEE 802.15.4 \cite{IEEE802.15.4} standard which is in an amendment in IEEE 802.15.4e-2015 \cite{IEEE802.15.4e}. 

\begin{figure}
    \centering
    \includegraphics[scale=.7]{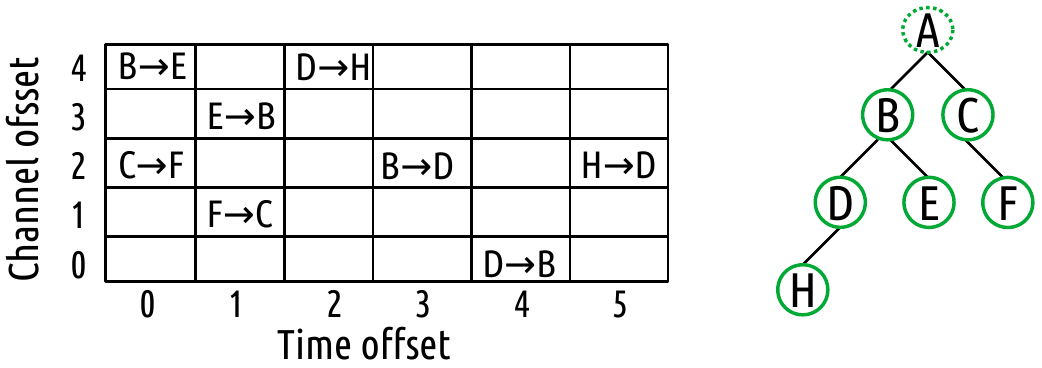}
    \caption{TSCH slotframe on the left and routing topology on the right}
    \label{fig:TSCH}
\end{figure}

TSCH is a link layer protocol which utilizes frequency hopping and synchronization techniques. Essentially neighboring nodes can transmit and receive packets based on a scheduler which defines the time slot and the channel. In order to avoid collisions it should not be allowed multiple packet transmission at the same time and channel. Thus, the schedule is organized in a two dimensional table called \textit{slotframe}. Figure \ref{fig:TSCH} depicts a slotframe along with the assumed topology. The $x$ axis in the slotframe stands for the time offset and the $y$ axis stands for the channel offset. A \textit{cell} corresponding to a specific time and channel offset is also called \textit{timeslot} and usually has a duration of $10\:ms$. In the depicted example of Figure \ref{fig:TSCH}, there are 6 timeslots and 5 channel offsets. The length of the slotframe is defined from the number of the timeslots and it is associated with several tradeoffs concerning the network performance. In each cell every node is allowed to perform one of the following actions, mandated by the schedule: (1) transmit a packet; (2) receive a packet; or (3) put the radio on sleep mode. For instance, in the mentioned example node B can send a packet to node E at the cell (0,4), node C can send a packet to node F at the cell (0,2) and the rest actions can be executed as defined from the schedule.

The TSCH coordinator is responsible for the construction of a TSCH network. More specifically, it defines the network ID, slotframe size and the Frequency Hopping Sequence (FHS). It also initializes the Absolute Slot Number (ASN) to 0 which is increased with each new timeslot. In order to join the TSCH network a node has to receive an Enhanced Beacon (EB) packet which is broadcast by the coordinator or other joined nodes. EB contains important information required like the time source to synchronize with the other nodes of the network. A channel for a given cell can be computed through Equation~\ref{eq:channel} where CO stands for Channel Offset. If there is more than one transmission cell at the same time, there is a priority mechanism for the cell with more packets.

\begin{equation}
    Channel = FHS(ASN + CO)\; mod\; FHS
\label{eq:channel}
\end{equation}

\subsection{Minimal Scheduling Function (MSF)}
\label{subsec:MSF}

The RFC 9033 \cite{rfc4180} defines a scheduling mechanism for TSCH implemented on the top of 6top protocol. With MSF scheduler every node defines some cells autonomously. An Rx cell is defined for the node itself using a hash function to decide the timeslot and the channel offset. The other cells of a slotframe are defined as Tx and are defined based on the traffic demands and they can be deleted if there is no traffic with a neighbor. 

\subsection{Orchestra}
\label{subsec:Orchestra}

Orchestra \cite{Duquennoy2015} was introduced in 2015. It is an autonomous scheduler for TSCH. One of the basic principles of Orchestra is that it combines different slotframes for MAC, routing and application purposes. Orchestra includes several types of operation. The Receiver Based (RB) and Sender Based (SB) types are the most popular ones which depend on storing the routing information to schedule unicast cells from parent to child. Each node is allocating a cell for its own (in RB type the cell is Rx and in SB Tx). The rest cells are Tx and Rx accordingly. A hash function using the neighbor address is constructing the schedule. For the evaluation purposes we use the RB type of operation. Every Orchestra timeslot includes two more slotframes, one for the default timeslot which is shared among all nodes and another one for EB packets which is broadcast either unicast. 

\subsection{Alice}
\label{subsec:Alice}

Alice \cite{Kim2019} is based on Orchestra but instead of calculating a schedule based on the node perspective it calculates it based on a link basis. Thus, it deploys a link-based channel offset instead of regular channel offset. Moreover, the unicast cells are updated once in every slotframe, based a hash function which is time dependent. 

\section{Related Work}
\label{sec:RelatedWork}

This section presents the state of the art of TSCH operating in mobile scenarios. Unfortunately, there are very few instances focusing on this topic, thus we consider it as motivation to investigate further this topic in order to provide insight and new solutions to these research questions.

Al-Nidawi et al. in \cite{AlNidawi2015} investigate the performance of TSCH and LLDN protocols under mobility scenarios. More specifically the authors focus on the overhead that joining and leaving the networks cause in terms of delay and energy. The authors claim that if a slotframe has a large channel offset then the scanning procedure will take more time and cost more energy since there are more channels to scan. This issue can be counter-measured with an according timeslot and scheduling approach. In addition, the authors show that TSCH can operate in mobile scenarios and do not consume more power if there is a sufficient number of nodes to provide a good network coverage. In a follow up paper \cite{Nidawi_b}, the same authors introduce Mobile TSCH, which is designed to reduce the delay which occurs when joining a TSCH network. The authors take advantage of ACK packets which are transmitted on a specific channel to advertise the TSCH network. In other words, the ACK packets behave as an Enhanced Beacon (EB) packet as well. The EB packets in TSCH are used to advertise the TSCH network and a node requires to successfully receive one in order to join the network. Thus, the scanning procedure takes place in a single channel and the delay and the energy consumption gets decreased.

A more recent attempt to investigate TSCH mobility is described in \cite{Raza2019}. The authors focus on cases where mobile nodes are leaving and re-joining a TSCH network due to issues which are related with the synchronization degree. They carry out an evaluation based on simulations which takes into account a combination of stationary and mobile nodes and measuring the downtime (the time a node is out of TSCH network), energy consumption, and the amount of Keep Alive (KA) packets (KA packets are used to update the synchronization of a node). Moreover, they regulate the number of the nodes and the speed, to see their impact on TSCH performance. The results describe a tradeoff between the downtime and the consuming energy connected with the operating speed. Essentially an increased speed might decrease the downtime but it will increase the energy consumption. Among other results they present that when the coverage of the network is sufficient with adequate number of mobile nodes, TSCH performance is very close to a static TSCH network.

The aforementioned approaches describe TSCH performance based on very basic mobility patterns and do not consider the state of the art of the TSCH schedulers. Furthermore these approaches consider a single mobility pattern in their evaluation method, but as it mentioned in Section \ref{sec:MobilityPatternsAndApplications}, mobility in IoT can vary a lot as it is dictated from several spatio-temporal variables. In contrast, the results presented in this paper are obtained considering the latest TSCH schedulers and we consider two mobile scenarios which differ in terms of randomness between them. Furthermore, the mobility patterns are inspired and designed based on real life application scenarios, thus the results reflect more to these cases.

\section{Mobility Patterns and Applications}
\label{sec:MobilityPatternsAndApplications}

This section describes the details of the mobility patterns used and the motivation behind it. The mobility pattern among IoT applications scenario vary a lot and includes several spatio-temporal variables which are difficult to be modelled in a general context. To this end we defined two scenarios which we focus on. The first one represents an application where autonomous agriculture vehicles operate on the field and the nodes follow a specific trail. The nodes are supposed to be part of a fleet collaboration system, enabling and allowing multiple autonomous agricultural robots to collaborate with each other as well as human operated tractors and machinery. TSCH is envisioned as part of a multiple-protocol solution with robust safety degradation features.

In the second one, a random movement mobility pattern represents moving machinery, autonomous robots/vehicles and workers who communicate in a smart warehouse environment. The patterns were implemented in Python and then imported in the Cooja simulator for the evaluation part. For the random movements in the smart warehouse pattern, we used the random library from Python, with a uniform distribution.

\subsection{Autonomous Agricultural Vehicle Mobility Pattern}

\begin{figure}
    \centering
    \includegraphics[scale=0.40]{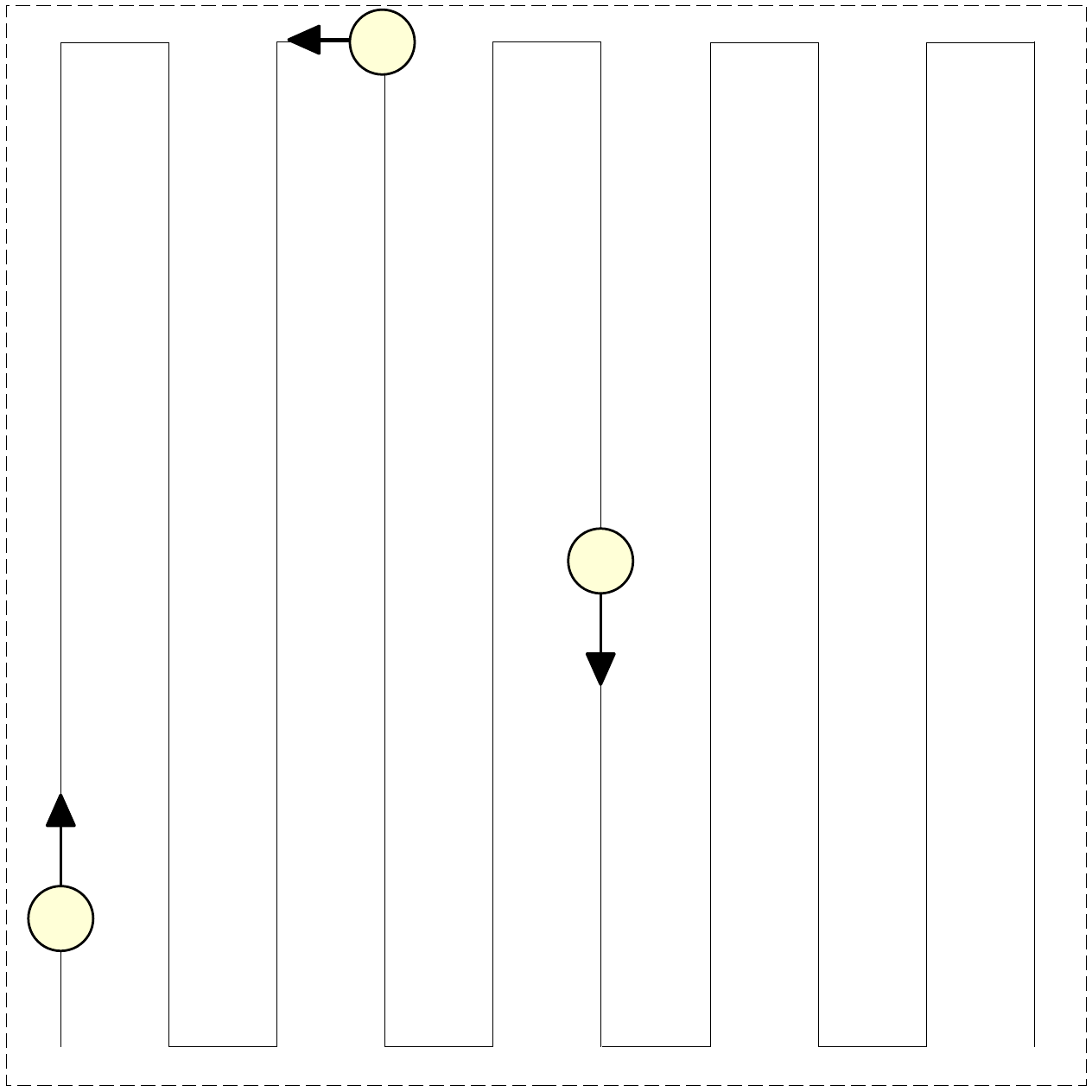}
    \caption{The mobility pattern which represents the function of autonomous agricultural vehicles operating on a field}
    \label{fig:ARF_mobility_pattern}
\end{figure}

In this mobility pattern the nodes move in a $1000\; $x\;$ 1000\; m$ frame in a predefined trail as it shown in Figure \ref{fig:ARF_mobility_pattern}. A node can start at any point of the predefined trail and move at any direction. When it will reach at the end it will start moving to the opposite direction. We assume that when two nodes are moving against each other they keep a safe distance between them. The pattern will continue till the end of the simulation. In a $4$-hour simulation which we execute, the nodes are repeating the trail $20$ times approximately. The purpose here is to check if the different TSCH schedulers are able to cope with the dynamic links which are created from such mobility pattern. This particular mobility pattern can be described as more deterministic from the sense that the nodes follow a repeating pattern several times.

\subsection{Smart Warehouse Mobility Pattern}

The nodes in this mobility pattern move in a same frame of $1000\; $x\;$ 1000\; m$ and it can start from any position. Then there are some limitations on which directions a node can move, in order to stay within the defined frame. For example, when a node is the upper left corner the available next moves are either right or down. Accordingly when a node is on the limit of the left side for instance, the available moves are right, down and up. If the node is at any other position it can move up, down right or left. The probabilities to choose a direction each time are $50\%$ for the first case, where there are two available directions to follow, $33.3\%$ for the second case, where the available options are three and $25\%$ for the last case where the directions a node can follow is four. It was used a uniform distribution in order to have an equal probability for each direction every time. The distance a node is moving every time is from $50$ to $500m$ with steps of 10 with a probability of $2.22\%$ for selecting each resulting distance. With this pattern we intended to represent how a node will move in a warehouse, with different movements every time and create a more random pattern to compare how the TSCH schedulers are able to perform on this scenario.
\section{Evaluation}
\label{sec:Evaluation}
This section presents the details of the evaluation method we used, the results we obtained and a discussion about the findings. 

To simulate the proposed scenarios, we used Cooja, \cite{cooja} a popular network simulator which is part of Contiki-OS \cite{Dunkels2004}. To enable mobility we also used a mobility plugin for Cooja \cite{mobility}. We first created several trails in Python following the mobility patterns described in Section \ref{sec:MobilityPatternsAndApplications}. In this sense every node will follow its own trail following a specific mobility pattern. The field a node can move is $1000\; $x\;$ 1000\; m$, the transmission range is $450\; m$ and the mobility speed $2\; m/s$. The simulation was based on Cooja motes, a virtual platform offered by the simulator which is able to execute Contiki-OS as a native process and utilize directly all the hardware accesses. The duration time of each simulation was $4$ hours. The slotframe size and the channel offset for the MSF scheduler is defined using a hash function which takes into account the amount of neighbors. Orchestra in our evaluation has a size of $397$ slots for EB slotframe, $31$ for the broadcast slotframe and for the Unicast slotframe it is constructed by the same hash function described before. For Alice the EB and Broadcast slotframe sizes are the same the Unicast though is constructed again every time the slotframe is completed based on directional link information and multiple channel offsets. 

We focus on a scenario where the traffic is directed from nodes to the coordinator. In this scenario every node is transmitting a packet every $6$ seconds. To have statistically valid results we executed the same simulation with $20$ different seeds for each selected scheduler and as we changed the number of nodes to observe how it can affect the performance of TSCH in each one. We conducted 360 simulations in total which correspond to $1440$ simulation hours and more than $3161000$ packets which were transmitted. However the initial positions of the nodes remain the same in order to investigate systematically when coverage issues that may or may not occur. For instance, the mobility pattern representing the autonomous agricultural vehicles include three to five nodes. In every simulation for the autonomous agricultural vehicles case, there is a node with severe coverage issues as it starts its trail away from the rest of the nodes and is having a hard time to join the network due to the fact that there is an increased time delay to join TSCH network and the time this node interacts with the rest of the nodes is not enough to complete this process. This is not happening in the smart warehouse mobility pattern, where none of the nodes spent significant time outside of the network. 

\subsection{Packet Received Ratio}
\label{subsection:PRR}

\begin{figure}
    \centering
    \includegraphics[scale=0.56]{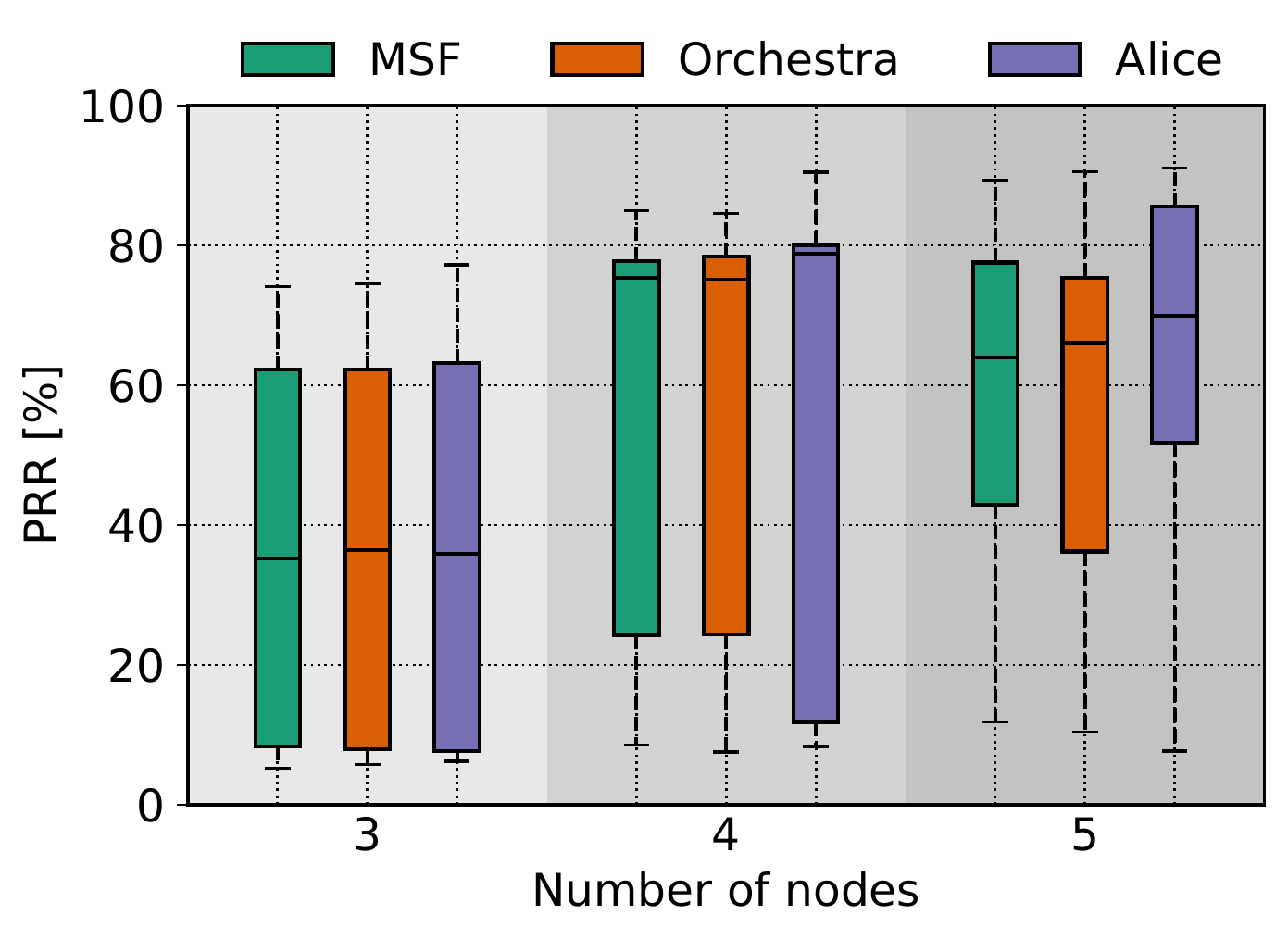}
    \caption{PRR for different TSCH schedulers under the autonomous agricultural vehicles mobility pattern, for different number of nodes.
    }
    \label{fig:PRR_ARF}
\end{figure}

\begin{figure}
    \centering
    \includegraphics[scale=0.56]{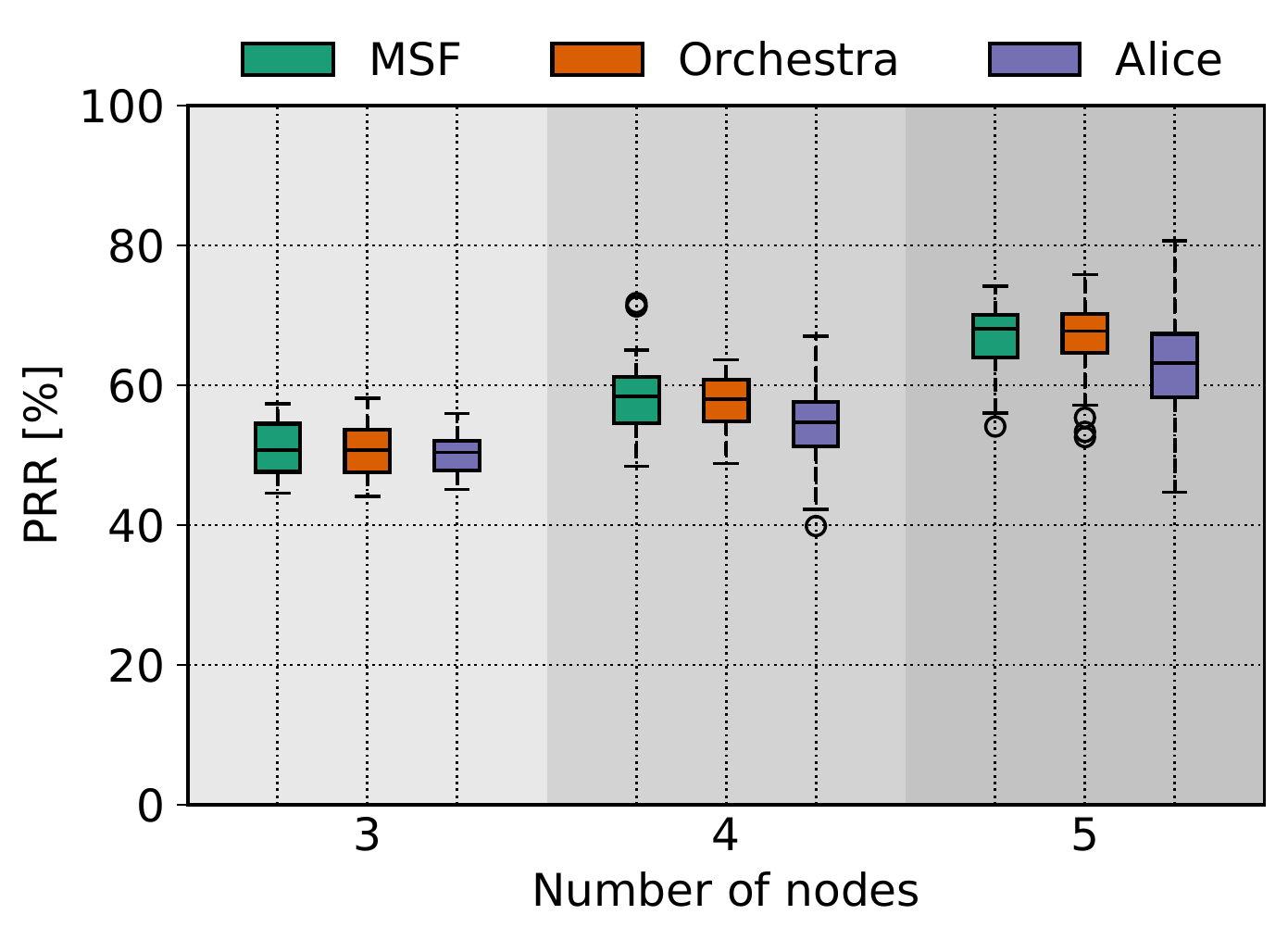}
    \caption{PRR for different TSCH schedulers under the smart warehouse mobility pattern, for different number of nodes.}
    \label{fig:PRR_Warehouse}
\end{figure}

The first metric we explore is the PRR which is the number of transmitted packets to the number of successfully received, including the retransmissions. Figures \ref{fig:PRR_ARF} and \ref{fig:PRR_Warehouse} represent the obtained results for the two mobility patterns accordingly. The results are presented with box plots figures where we group the TSCH scheduling approaches as we differentiated the amount of operating nodes in the x axis. 

We observe the high variance in Figure \ref{fig:PRR_ARF} where we evaluate the autonomous agricultural vehicles scenario. This is because most of the nodes achieve similar amount of PRR but the one which starts from a remote position is having a significant lower value and the boxes in the plot represent the values of all nodes in the network. 

\begin{figure}
    \centering
    \includegraphics[scale=0.56]{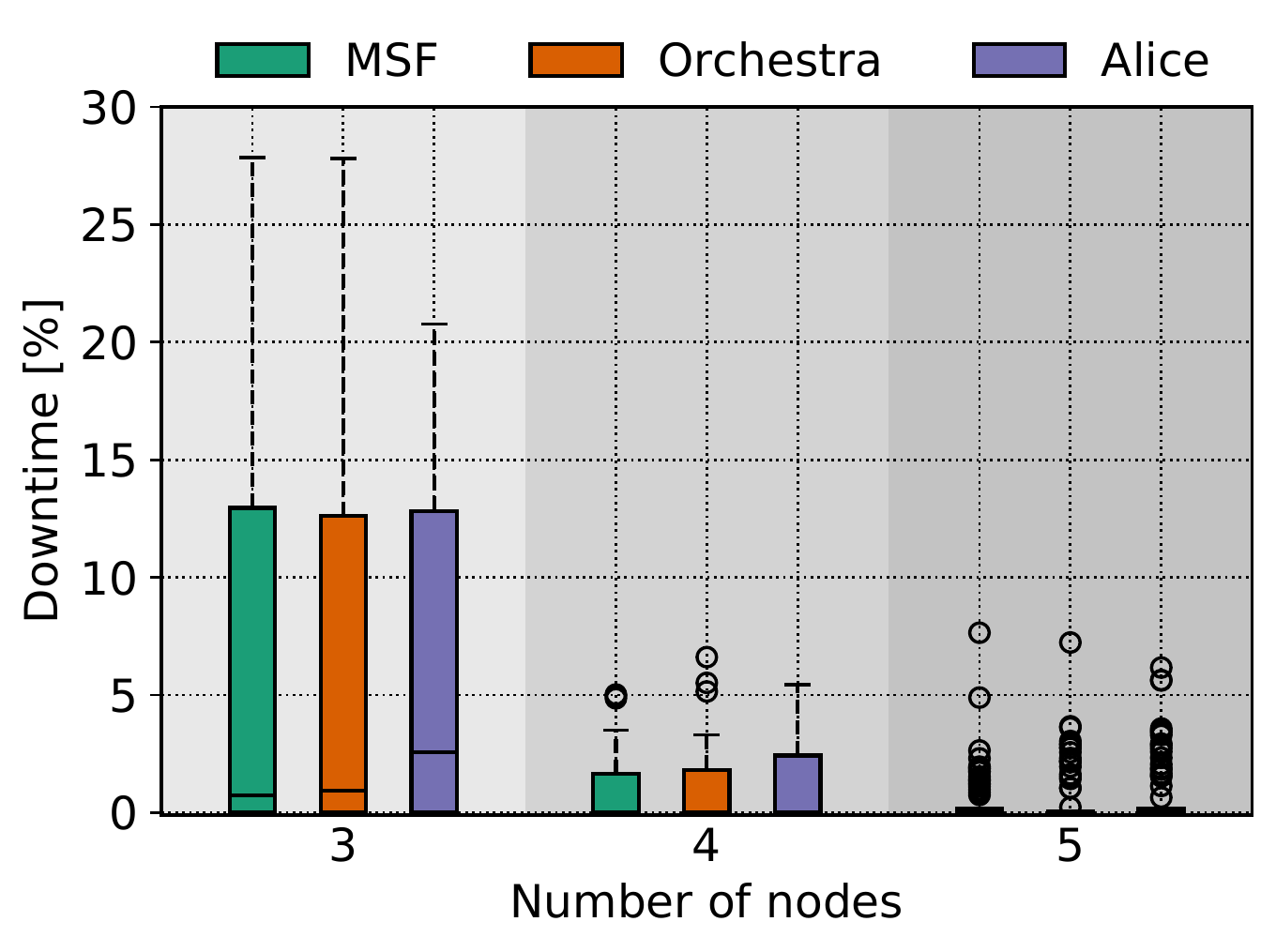}
    \caption{Downtime for different TSCH schedulers under the autonomous agricultural vehicles mobility pattern, for different number of nodes. The amount of downtime is the percentage of a 4 hour simulation.}
    \label{fig:Downtime_ARF}
\end{figure}

The next observation has to do with the number of operating nodes. One might have assumed that if you increase the number of nodes the coverage would increase and therefore the PRR would be higher \cite{Raza2019}. While we believe this is a valid argument and this is a trend we see in Figures \ref{fig:PRR_ARF} and \ref{fig:PRR_Warehouse} we also observe that the PRR value decreases in Figure \ref{fig:PRR_ARF} when we increase the number of nodes from 4 to 5. This is not happening because one the new nodes has increased downtime value, which is confirmed by Figure \ref{fig:Downtime_ARF} but rather because of the TSCH mechanics and new traffic demands because of the new node joining the network. As downtime we define the time a node which is not associated with the TSCH network. This can happen either due to the fact that a node is out of range because of the mobility pattern but also due to synchronization or high traffic and scheduling issues. It is important to be highlighted that when TSCH is used in mobile scenarios the downtime is not only occurring out of coverage issues but also because of scheduling approaches and low synchronization issues. Apparently, these problems contribute to decrease the network performance as well.

Regarding the different schedulers we used, we do not observe any significant difference in the PRR. ALICE performs slightly better for the autonomous agricultural vehicle mobility pattern but for the case of the smart warehouse mobility pattern, ALICE performs slightly worse than the other two scheduling approaches. We speculate that the directional link approach that ALICE employs is more efficient for a more deterministic mobility pattern than a more stochastic one like the smart warehouse one.

Next if we compare the PRR between the two mobility patterns we will observe that there is a different performance and it is difficult to compare in a holistic manner. For instance, if we compare the PRR values for the case when 3 nodes were operating, in the autonomous agricultural vehicle case in Figure \ref{fig:PRR_ARF} the PRR values are very low because one of the three nodes is having significant downtime and very low PRR consequently compared with the ones in the smart warehouse case in Figure \ref{fig:PRR_Warehouse}. If we compare now the case where there are 4 nodes operating in Figure \ref{fig:PRR_ARF} we see that the PRR values are increased by a large degree because the new introduced node operates within the network range and its PRR value is able to compensate in the mean value depicted in the box plot. 

The main remark is that the PRR values are very low in general for all cases and the current TSCH approaches cannot provide a reliable and robust performance when they are used in a mobile scenario.
\subsection{Downtime}
\label{subsection:Downtime}

The downtime is a crucial metric which points out the time a node might spent not associated with TSCH network. As it was mentioned before, beside coverage issues other issues like going out of synchronization or high traffic might lead a node to leave the network. Being out of coverage is a problem that cannot be solved with a straight forward way but leaving the TSCH network because of synchronization issues or because of the scheduler does not include enough amount of EB cells during a mobile scenario looks more manageable problem.

\begin{figure}
    \centering
    \includegraphics[scale=0.56]{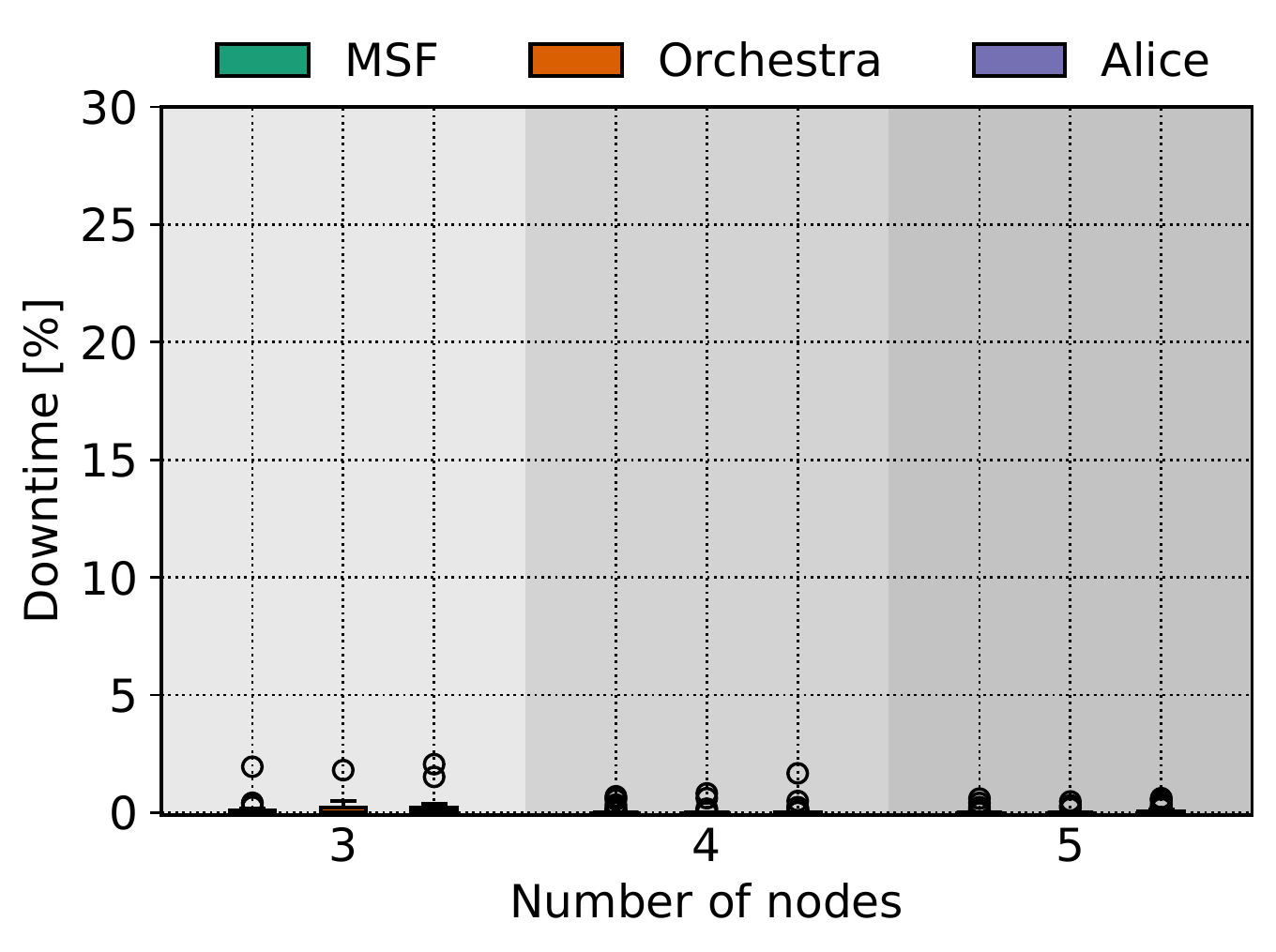}
    \caption{Downtime for different TSCH schedulers under the smart warehouse mobility pattern, for different number of nodes. The amount of downtime is the percentage of a 4 hour simulation.}
    \label{fig:Downtime_Warehouse}
\end{figure}

In the cases we investigated the downtime was rather low in general specifically for the smart warehouse mobility pattern as it is reported very close to 0 in Figure \ref{fig:Downtime_Warehouse}. For the autonomous agricultural vehicles case the downtime is increased in some cases since one of the nodes is having an increased downtime on purpose.

\subsection{Initial Join Time}
\label{subsection:Jointime}

\begin{figure}
    \centering
    \includegraphics[scale=0.56]{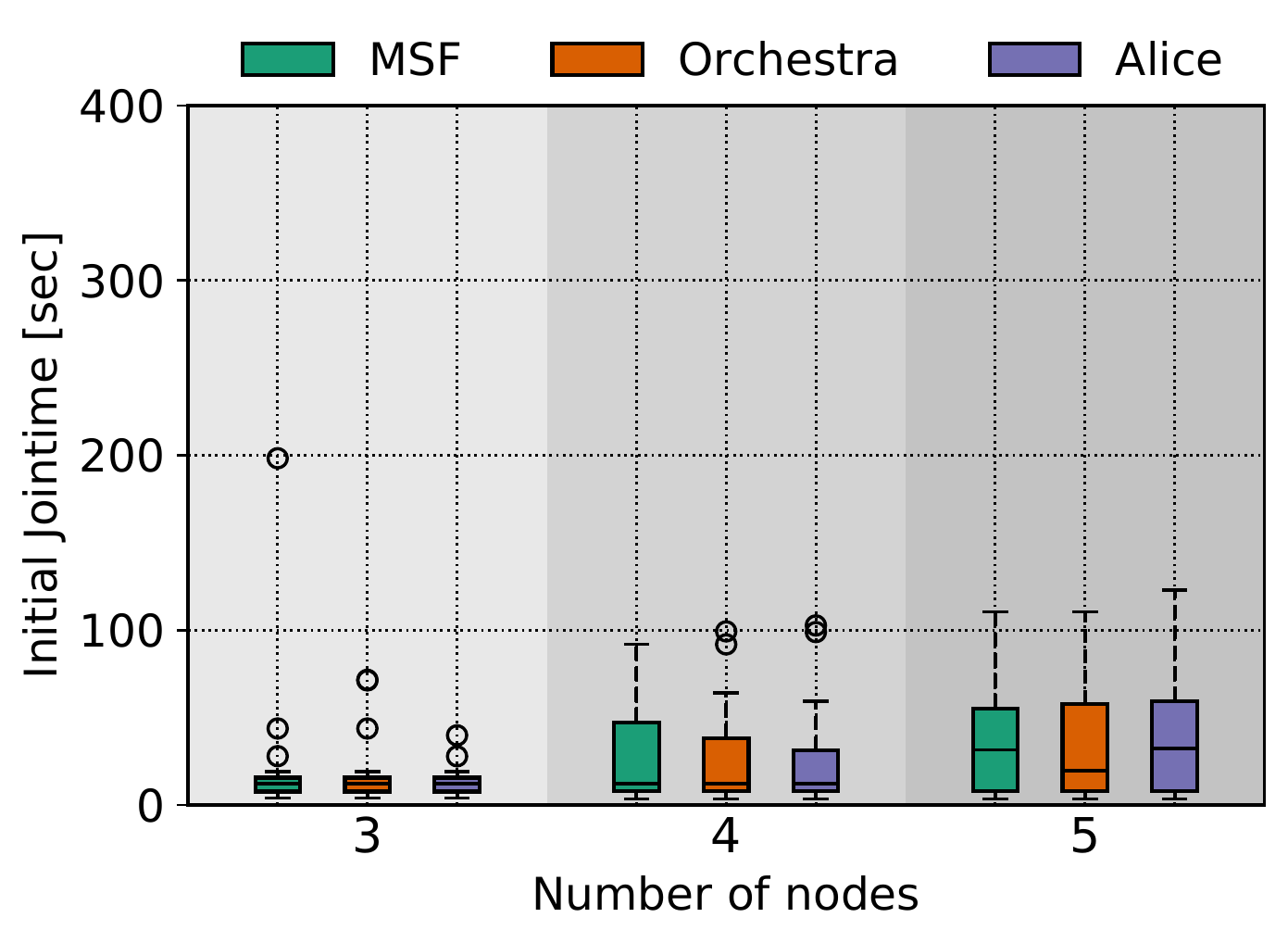}
    \caption{Jointime for different TSCH schedulers under the autonomous agricultural vehicles mobility pattern, for different number of nodes.}
    \label{fig:jointime_ARF}
\end{figure}

\begin{figure}
    \centering
    \includegraphics[scale=0.56]{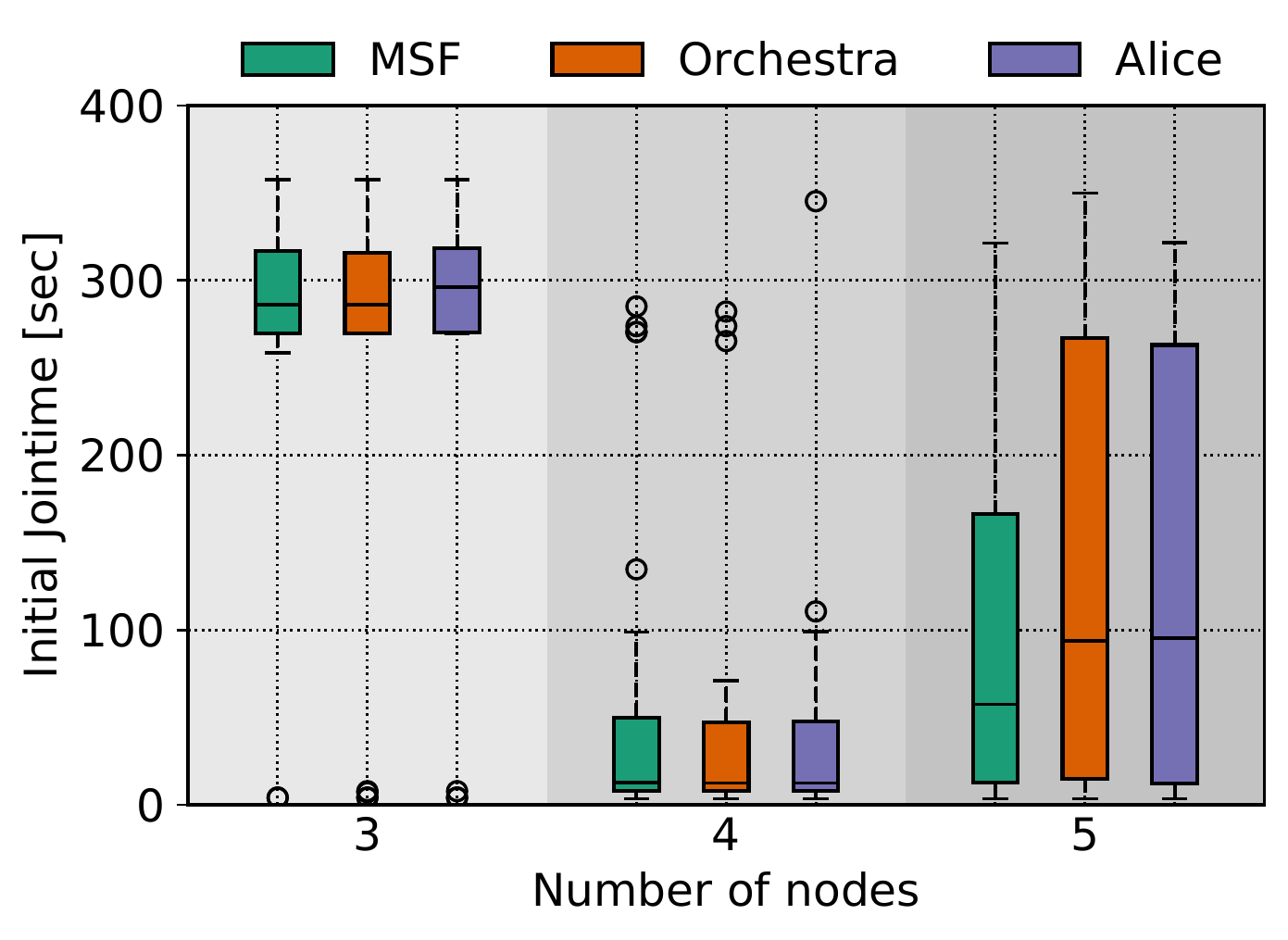}
    \caption{Jointime for different TSCH schedulers under the smart warehouse mobility pattern, for different number of nodes.}
    \label{fig:jointime_Warehouse}
\end{figure}

We also report the initial jointime metric, which is the amount of time it takes a node to join the TSCH network from the time it boots. This is an important metric since it is known that for some TSCH approaches it takes a notable amount of time to join the TSCH network. If we combine this with a mobile scenario the amount of time can be increased by a lot. Figures \ref{fig:jointime_ARF} and \ref{fig:jointime_Warehouse} illustrate the results. 

In Figure \ref{fig:jointime_Warehouse} we see that when the operating nodes are 3 the initial jointime is increased but when the nodes are increased  to 4, the jointime drops very close to 0. We speculate that in the second case the new node act as the mediator to provide coverage to the rest of the nodes in the beginning and helps them to associate with the TSCH network faster.

We have to mention also the case when a node will be deployed in an initial position which might be out of coverage regarding the rest nodes. The chances of this node to join the TSCH network is based on how much time it will interact with the rest of the nodes and if the joining time of the regarding TSCH approach is increased, the node will have fewer chances to join the network. To this end, it is important to have a short joining time because in cases like these, the chances this node to report the sensed data will be based on if it will manage to join the network during this short interval. 
\section{Conclusion}
\label{sec:Conclusion}

In this paper we report how TSCH schedulers perform during heterogeneous mobile scenarios and where we should focus to improve it for such cases. We defined two mobility patterns representing two different mobile applications which the first one is more deterministic, as it is consisted of repetitive patterns, and the second one is more stochastic. We run simulations after selecting TSCH approaches from the state of the art and perform evaluation terms of PRR, downtime and the initial join time. The results show that the current approaches cannot be utilized in mobile scenarios as their performance is inadequate. The main contribution of the paper is in the identification of the directions researchers should concentrate in order to improve TSCH performance towards mobile scenarios.

\bibliographystyle{abbrv}
\bibliography{references}

\begin{thebibliography}{10}

\bibitem{mobility}
{Mobility of Nodes in Cooja}.
\newblock
  \url{https://anrg.usc.edu/contiki/index.php/Mobility_of_Nodes_in_Cooja}.
\newblock Accessed: 2021-06-16.

\bibitem{IEEE802.15.4e}
{IEEE Standard for Local and metropolitan area networks--Part 15.4: Low-Rate
  Wireless Personal Area Networks (LR-WPANs)}.
\newblock {\em IEEE Std 802.15.4-2011 (Revision of IEEE Std 802.15.4-2006)},
  pages 1--314, 2011.

\bibitem{IEEE802.15.4}
{IEEE Standard for Low-Rate Wireless Networks}.
\newblock {\em IEEE Std 802.15.4-2015 (Revision of IEEE Std 802.15.4-2011)},
  pages 1--709, 2016.

\bibitem{Nidawi_b}
Y.~Al-Nidawi and A.~H. Kemp.
\newblock {Mobility Aware Framework for Timeslotted Channel Hopping IEEE
  802.15.4e Sensor Networks}.
\newblock {\em IEEE Sensors Journal}, 15(12):7112--7125, 2015.

\bibitem{AlNidawi2015}
Y.~Al-Nidawi, H.~Yahya, and A.~H. Kemp.
\newblock {Impact of mobility on the IoT MAC infrastructure: IEEE 802.15.4e
  TSCH and LLDN platform}.
\newblock In {\em 2015 IEEE 2nd World Forum on Internet of Things (WF-IoT)},
  pages 478--483, 2015.

\bibitem{BRUNLAGUNA201883}
K.~Brun-Laguna, A.~L. Diedrichs, D.~Dujovne, C.~Taffernaberry, R.~Léone,
  X.~Vilajosana, and T.~Watteyne.
\newblock {Using SmartMesh IP in Smart Agriculture and Smart Building
  applications}.
\newblock {\em Computer Communications}, 121:83--90, 2018.

\bibitem{rfc4180}
T.~Chang, M.~Vučinić, X.~Vilajosana, S.~Duquennoy, and D.~Dujovne.
\newblock {6TiSCH Minimal Scheduling Function (MSF)}.
\newblock RFC 9033, 05 2021.

\bibitem{WirelessHART}
D.~Chen, M.~Nixon, and A.~Mok.
\newblock {\em WirelessHART(TM): Real-Time Mesh Network for Industrial
  Automation}.
\newblock Springer, 2010.

\bibitem{Guglielmo2016}
D.~De~Guglielmo, B.~Al~Nahas, S.~Duquennoy, T.~Voigt, and G.~Anastasi.
\newblock {Analysis and Experimental Evaluation of IEEE 802.15.4e TSCH CSMA-CA
  Algorithm}.
\newblock {\em IEEE Transactions on Vehicular Technology}, 66(2):1573--1588,
  2017.

\bibitem{Dunkels2004}
A.~Dunkels, B.~Gronvall, and T.~Voigt.
\newblock Contiki - a lightweight and flexible operating system for tiny
  networked sensors.
\newblock In {\em 29th Annual IEEE International Conference on Local Computer
  Networks}, pages 455--462, 2004.

\bibitem{Duquennoy2015}
S.~Duquennoy, B.~Al~Nahas, O.~Landsiedel, and T.~Watteyne.
\newblock {Orchestra: Robust Mesh Networks Through Autonomously Scheduled
  TSCH}.
\newblock In {\em Proceedings of the 13th ACM Conference on Embedded Networked
  Sensor Systems}, SenSys '15, page 337–350, New York, NY, USA, 2015.
  Association for Computing Machinery.

\bibitem{Duquennoy2017}
S.~Duquennoy, A.~Elsts, B.~A. Nahas, and G.~Oikonomou.
\newblock {TSCH and 6TiSCH for Contiki: Challenges, Design and Evaluation}.
\newblock In {\em 2017 13th International Conference on Distributed Computing
  in Sensor Systems (DCOSS)}, pages 11--18, 2017.

\bibitem{Elsts18}
A.~Elsts, X.~Fafoutis, P.~Woznowski, E.~Tonkin, G.~Oikonomou, R.~Piechocki, and
  I.~Craddock.
\newblock {Enabling Healthcare in Smart Homes: The SPHERE IoT Network
  Infrastructure}.
\newblock {\em IEEE Communications Magazine}, 56(12):164--170, 2018.

\bibitem{Hamideh2019}
H.~Hajizadeh, M.~Nabi, R.~Tavakoli, and K.~Goossens.
\newblock {A Scalable and Fast Model for Performance Analysis of IEEE 802.15.4
  TSCH Networks}.
\newblock In {\em 2019 IEEE 30th Annual International Symposium on Personal,
  Indoor and Mobile Radio Communications (PIMRC)}, pages 1--7, 2019.

\bibitem{Kim2019}
S.~Kim, H.-S. Kim, and C.~Kim.
\newblock {ALICE: Autonomous Link-based Cell Scheduling for TSCH}.
\newblock In {\em 2019 18th ACM/IEEE International Conference on Information
  Processing in Sensor Networks (IPSN)}, pages 121--132, 2019.

\bibitem{Xiulong2018}
X.~Liu, J.~Cao, Y.~Yang, and S.~Jiang.
\newblock {CPS-Based Smart Warehouse for Industry 4.0: A Survey of the
  Underlying Technologies}.
\newblock {\em Computers}, 7(1), 2018.

\bibitem{cooja}
F.~Osterlind, A.~Dunkels, J.~Eriksson, N.~Finne, and T.~Voigt.
\newblock {Cross-Level Sensor Network Simulation with COOJA}.
\newblock In {\em Proceedings. 2006 31st IEEE Conference on Local Computer
  Networks}, pages 641--648, 2006.

\bibitem{Bigi2019}
B.~V. Philip, T.~Alpcan, J.~Jin, and M.~Palaniswami.
\newblock {Distributed Real-Time IoT for Autonomous Vehicles}.
\newblock {\em IEEE Transactions on Industrial Informatics}, 15(2):1131--1140,
  2019.

\bibitem{Raza2019}
S.~Raza, T.~v.~d. Lee, G.~Exarchakos, and M.~Güneş.
\newblock {A Reliability Analysis of TSCH Protocol in a Mobile Scenario}.
\newblock In {\em 2019 16th IEEE Annual Consumer Communications Networking
  Conference (CCNC)}, pages 1--6, 2019.

\bibitem{Vilajosana2014}
X.~Vilajosana, Q.~Wang, F.~Chraim, T.~Watteyne, T.~Chang, and K.~S.~J. Pister.
\newblock {A Realistic Energy Consumption Model for TSCH Networks}.
\newblock {\em IEEE Sensors Journal}, 14(2):482--489, 2014.

\bibitem{Watteyne2016}
T.~Watteyne, A.~L. Diedrichs, K.~Brun-Laguna, J.~E. Chaar, D.~Dujovne, J.~C.
  Taffernaberry, and G.~Mercado.
\newblock {PEACH: Predicting Frost Events in Peach Orchards Using IoT
  Technology}.
\newblock {\em EAI Endorsed Transactions on Internet of Things}, 2(5), 12 2016.

\end{thebibliography}

\end{document}